\documentclass[conference]{IEEEtran}
\usepackage{cite}
\usepackage{amsmath,amssymb,amsfonts}
\usepackage{algorithmic}
\pdfoutput=1
\usepackage{subfig}
\usepackage{graphicx}
\usepackage{textcomp}
\usepackage{xcolor}
\usepackage{booktabs}
\usepackage{multicol}
\usepackage{lipsum}
\usepackage{tikz}
\usepackage{capt-of}
\usepackage{float} 
\usepackage{comment} 
\usepackage{amsmath}
\usetikzlibrary{shapes.misc}
\tikzset{cross/.style={cross out, draw=black, minimum size=3*(#1-\pgflinewidth), inner sep=0pt, outer sep=0pt},
cross/.default={3pt}}

\def\BibTeX{{\rm B\kern-.05em{\sc i\kern-.025em b}\kern-.08em
    T\kern-.1667em\lower.7ex\hbox{E}\kern-.125emX}}
    
\begin{document}

\title{Estimating Phase Aberration from Noisy Radiofrequency Data of a single frame of Synthetic Aperture Ultrasound Image}

\author{\IEEEauthorblockN{1\textsuperscript{st} Dena Monjazebi}
\IEEEauthorblockA{\textit{dept. Physics } \\
\textit{Ryerson University}\\
Toronto, Canada \\
dena.monjazebi@ryerson.ca}
\and
\IEEEauthorblockN{2\textsuperscript{nd} Yuan Xu}
\IEEEauthorblockA{\textit{dept. Physics} \\
\textit{Ryerson University}\\
Toronto, Canada \\
yxu@ryerson.ca}
}

\thanks{Dena Monjazebi was with the Department
of Science, Ryerson University, Toronto,
ON, Canada e-mail: dena.monjazebi@ryerson.ca.}

\maketitle

\begin{abstract}
Phase aberration is one of the main contributors to image degradation in ultrasound imaging.  Normalized-Cross-Correlation (NCC) is one of the most extensively studied techniques to estimate the arrival delay error and the aberration profile. \textit{However, the performance of NCC can be compromised when the data have a poor signal-to-noise-ratio.} Here we propose an iterative filtered NCC (f-NCC) method to estimate phase aberration in speckle regions from the noisy radio-frequency data of a single frame of synthetic transmit aperture (STA) image. First, two-dimensional filters were applied in the aperture and temporal domains to suppress noise and off-axis signals in the transmit beamformed data from an arbitrary speckle region. Second, an  iterative method that included the phase aberration in both the transmission and receiving process was used to estimate the phase aberration. Third, the estimation were improved by correcting the edge effect with an iterative zero-padding technique. Last, in the experimental study, we proposed and validated a formula to estimate the arrival time error caused by a sound speed error. The results from both the numerical simulations and experimental studies confirm that the proposed filtered NCC method can estimate the phase aberration profile correctly from noisy STA RF data in a speckle region. Image quality in terms of contrast, resolution and image signal-to-noise-ratio was improved. Potential applications of the proposed method to map local speed of samples are also discussed.

\end{abstract}
\section{Introduction}

\IEEEPARstart{P}{ha}se aberration can compromise the performance of both ultrasound imaging and ultrasound therapy. Recent studies has shown that phase aberration due to sound speed heterogeneity is the dominant cause of image degradation resulted from poor beam-focusing quality\cite{Tabei2003,Guenther2009,Mann2002,Bell2015,Chau2017}. In the presence of speed heterogeneity, delay calculations solely based on the geometric analysis are incorrect. These errors will cause poor focusing, reduce contrast and worsen lateral resolution. With fatty tissue having a different sound-speed compared to other organs, phase aberration is a major problem in ultrasound imaging of breast and obese patients.

Phase aberration should be compensated to align the wavefront during beamforming. Many approaches have been developed to correct the phase aberration. Some of these methods are: normalized cross-correlation (NCC) \cite{ODonnell1988,Krishnan1996,ODonnell1988a,Liu1994,Ng1994}, maximization of speckle brightness \cite{Nock1989}, time reversal of ultrasound field \cite{Fink1992} and minimization of the sum of absolute differences between signals at adjacent elements \cite{Karaman1993}. Aperture domain model image reconstruction (ADMIRE) has been proven effective to reduce the acoustic clutter by decomposing the aperture domain data to differentiate between clutter and main beam \cite{Byram2015}.

NCC is a widely studied delay estimation algorithm in ultrasound imaging and therapy. O'Donnell and Flax have demonstrated that normalized and non-normalized cross-correlation techniques can be used to obtain accurate estimations of delay compensation in B-mode imaging \cite{ODonnell1988}\cite{Ng1994}. Generally, a segment of radiofrequency (RF) data in a reference received signal was compared with a segment of RF data in a delayed received signal. A pattern-matching function was utilized to find the delay that maximized the resemblance of the two segments \cite{Shaswary2015}. Regarding commonly used algorithms, normalized and non-normalized NCC gave reasonable results in terms of precision and computational time \cite{Viola2003}. Many papers have used the NCC aberration profiles information to estimate an average speed of the medium or a local speed map \cite{Byram2012,Anderson1998,Jakovljevic2018}. 

Efficacy of NCC studies highly depends on the Signal-to-Noise-Ratio(SNR) of the data. This requirement introduces a major shortcoming in \textit{in vivo} studies. In hard-to-image patients due to the extra intermediate fat layer, organs and anatomical features are located at a larger depth. With the additional layer of fat attenuating the signal, RF signals from organs of interest have inherently lower SNR. Slack et al. have indicated prevalence of obesity in US as 30.8 \%. This Study reported a possible projection of 42 \% by 2030 \cite{Slack2014}. In clinical abdominal ultrasound, over-weightness is the biggest limitation to produce optimal quality images\cite{Alm2008}. The requirement for high SNR signals, introduces a major obstacle to expand the clutter suppression methods into clinic. With the rising rates of obesity, the need for a successful phase aberration correction method for low $RF_{SNR}$ in ultrasound visualisation is increasing. 

Various methods with the goal to correct the dilemma of phase aberration correction method in presence of noise has been developed. Many aperture domain filters have been investigated to reduce the off-axis signals and noise \cite{Dahl2003,Dahl2008}. Shin et al. have proposed a method with the focus on phase aberration correction for RF in presence of of noise \cite{Shin2018}. Other methods include spatial compounding to suppress phase aberration and noise clutters together\cite{Dahl2017}. 

Most phase aberration methods were discussed for B-mode imaging. In this paper, we focus on synthetic transmit aperture (STA) image due to feasibility of using the RF data of a single frame of STA image to estimate the phase aberration iteratively. In STA data acquisition, a semi-spherical wave is sent by a single transmission element to cover a large imaging area. The echo signals are recorded by the entire receive aperture. This received data can be used to reconstruct a low-resolution image. By repeating the transmission across the entire aperture, a three-dimensional STA data set can be acquired to reconstruct a high-resolution image. STA imaging can achieve the optimum transmission and receiving focus across the whole image. In addition, STA data set has more information than that of B-mode imaging \cite{Gong2015}.

However, it is known that STA RF data exhibits a smaller $RF_{SNR}$ due to smaller transmission energy than the conventional B-mode data. In this paper, we propose a solution to improve the NCC method for noisy RF signals by filtering the RF data in the aperture-temporal domain. A low-pass aperture-temporal domain filter was designed to pass the main beam signals and filter out the acoustic clutter and noise. Previous research indicated off-axis scattering can be reduced by a aperture-domain filter to improve the phase aberration estimations in B-mode imaging\cite{Dahl2003,Dahl2008}. \textit{Here, f-NCC method was developed for areas with speckles in STA imaging where $RF_{SNR}$ is significantly lower than B-mode.}

In the method section, we will introduce the low-pass aperture-temporal domain filter that applies gradual cut-offs in aperture and temporal frequency domains. The efficacy of this filter to minimize the clutter and off-axis scattering in the transmit-beamformed (TBF) signals will be investigated. In the results section, the f-NCC algorithm will be studied in simulation and experimental level. Alternative potential of f-NCC methodology and future work will be explained in the discussion section.  Finally the results and conclusions will be summarized in the conclusion section. The appendices include an introduction to NCC, zero-padding to alleviate the edge effect, and the derivation of the phase aberration profile in a average speed error scenario.    

\section{METHOD}
\subsection{Iterative estimation of phase aberration in STA}
In this study, we utilized an NCC-based iterative method to estimate the phase aberration \cite{Karaman1993} in Synthetic Transmit Aperture (STA) ultrasound imaging since it does not require multiple rounds of data acquisition. Let us first define a complete set of STA RF data: $P_{ij}(t)$, which was the signal at time $t$ from the $i_{th}$ ($i = 1:N$) receive element upon the activation of the $j_{th}$ ($j = 1:M$) transmission element. We used \(g_{i,j}(x_f,z_f)\) to represent the focusing or geometrical delay corresponding to the focal point at \(x_f\) and \(z_f\). The geometrical delay was defined as the travel time from transmit element $j$ to the focal point and back to the receive element $i$. We used $s_i(t)$ to represent the transmission beamformed (TBF) RF signals. This represents the RF signals received by the $i_{th}$ ($i = 1:N$) receive element when the transmitted  beam was focused at the focal point (\(x_f\), \(z_f\)). 

TBF signal $s_i(t)$ was obtained by adopting the Delay-and-Sum (DAS) algorithm: summing over all transmit elements of the RF signals from the $i_{th}$ receive element (Eq. \ref{eq20}) after applying the geometrical delay $g_{ij}$ and the estimated phase aberration $\{t_i\}$ to the STA RF signals,

\begin{equation}\label{eq20}
s_i(t)= \sum_{j=1}^{128} {P_{ij}{(t-g_{i,j}(x_f,z_f)}+t_i+t_j)}.
\end{equation}

Notice that the phase aberration correction was applied to both $j_{th}$ transmit element and $i_{th}$ receive element. We assumed that each element has experienced the same phase aberration in both the transmission and receiving process.

A plot of $s_i(t)$ was shown in Fig. \ref{fig:sub1}(a). The focal point was at $x_f=0.96 cm$ and $z_f=1.63cm$ as signified by the red square in Fig. \ref{fig:sub1}(b). $s_i$ was the signal received at the $i_{th}$ receiving elements after considering the geometrical delay when a beam focused at \(x_f\) and \(z_f\) was transmitted. In this paper depth and time were used interchangeably. Additionally, we defined the echoes due to the scattering at the \(x_f\) and \(z_f\) as wavefronts. For example, we considered the signals within a window centered around the depth of 1.63 cm at Fig. \ref{fig:sub1}(a) as wavefronts.  When there is no phase aberration, the echoes from the scattering at \(x_f\) and \(z_f\) will make a straight line (see Fig. \ref{fig:sub1}(a) for an example). In reality, there is always phase aberration, so the wavefront will be distorted like Fig. \ref{fig:sub7}(a). Our goal was to find the proper delay error that could resolve this distortion. 
Here, we introduced a method to estimate the phase aberration iteratively \cite{Karaman1993} as follows. We started with an initial guess of phase aberration $\{t_i^0\}$ equal to zero, then obtain the TBF RF data based on the current $t_j^{k-1}$.

\begin{equation}\label{eq6}
s_i^{k}(t) = \sum_{j=1}^{128} {P_{ij}{(t-g_{i,j}(x_f,z_f)}+t_i^{k-1}+t_j^{k-1})}
\end{equation}

Here, $s_i^{k}(t)$ represented the corrected TBF RF data at the $k_{th}$ iteration. $t_i^{k-1}$ was the accumulated phase aberration estimated in the previous ($k-1$) iterations for the $i_{th}$ receive element. 
Next, NCC or f-NCC were applied to $s_i^{k}(t)$ to obtain the residual phase aberration $\{r_i^k\}$ in the $k_{th}$ iteration. See details of finding the residual phase aberration $\{ r^k_i\}$ in the appendix A. Then the $\{t_i^k\}$  was updated as:  $t_i^k= t_i^{k-1}+ r_i^k$ . After that the updated $\{t_i^k\}$  was applied to Eq. \ref{eq6} for the next iteration until $\{r_i^k\}$ converged to zero. 

\subsection{Filtered NCC}
One of the major shortcomings of NCC techniques is the inadequacy of conventional NCC method for RF signals with low SNR. As explained in the introduction section, this requirement can introduce a major problem in STA data in experimental and \textit{in-vivo} studies. Therefore it is worthwhile to investigate a solution. 

\begin{figure}
\centering
   \subfloat[][]{\includegraphics[width=.4\textwidth]{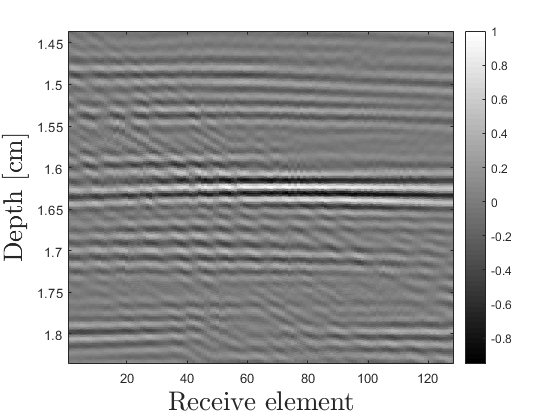}} \quad
   \subfloat[][]{\includegraphics[width=.4\textwidth]{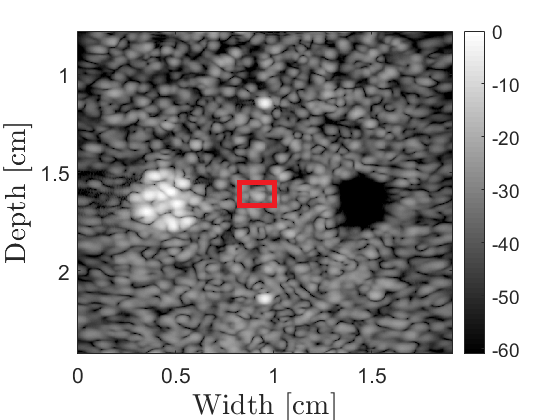}}
   \caption{ (a) plot of $s_i(t)$, transmit beamformed RF signals, corresponding to the speckle inside the red box in (b).  (b) the reconstructed image in the absence of noise and aberration. Focal depth was at 1.63cm.}
   \label{fig:sub1}
\end{figure}

In addition to the electronic noise, off-axis scattering can interfere with the wavefront of interest. For example, the oblique wavefront in Fig. \ref{fig:sub1}(a) was due to the scattering of the off-axis beams (side-lobes or grating lobes) from the hyper-echoic lesion on the left side of the speckle at $x=0.98$ cm and $z=1.63cm$ in Fig. \ref{fig:sub1}(b). This off-axis clutter interfered with the estimation of phase aberration. The inaccurate estimations by NCC illustrated in Fig. \ref{fig:sub4} are due to this interference. The goal of this study was to develop a phase aberration estimator in STA imaging in the presence of noise and clutter in RF data. We investigated solutions that would reduce noise and off-axis signal while preserving the shape of the wavefront. The aim is to consequently increase the SNR and Signal-clutter-ratio (SCR) of TBF RF signal, $s_i(t)$. \textbf{A successful filtering method should not change the temporal wavefront distortion caused by phase aberration. }

\begin{figure}
\centering
\begin{tikzpicture}

\node[inner sep=0pt] (org) at (0,0)
    {\includegraphics[width=.35\textwidth]{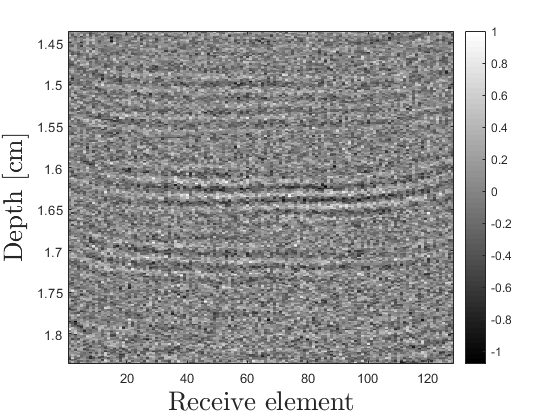}};
     \draw[fill=white] (-2,1.6) circle (0.28cm) node[text=black] {a};
      \draw[fill=white] (0,2.3)  node[text=black] {\large$s_i(t)$};
   
     \node[inner sep=0pt] (fft) at (10,0)
    {\includegraphics[width=.35\textwidth]{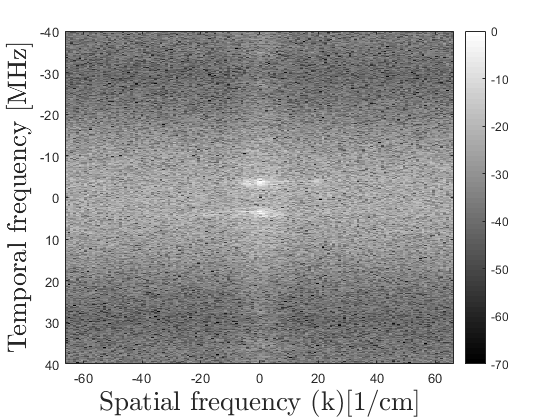}};
    \draw[fill=white] (8.0,1.6) circle (0.28cm) node[text=black] {b};
    \draw[fill=white] (10,2.3)  node[text=black] {\large$S_k(f)$};
            \draw[->,line width=0.8mm] (org.east) -- (fft.west)
    node[midway,fill=white] {\large 2D FFT};
        \draw (9.8,-2.65) node[cross,rotate=0] {};
        \draw (9.8,-2.65) node[cross,rotate=45] {};
   
     \node[inner sep=0pt] (filter) at (10,-5.5)
    {\includegraphics[width=.35\textwidth]{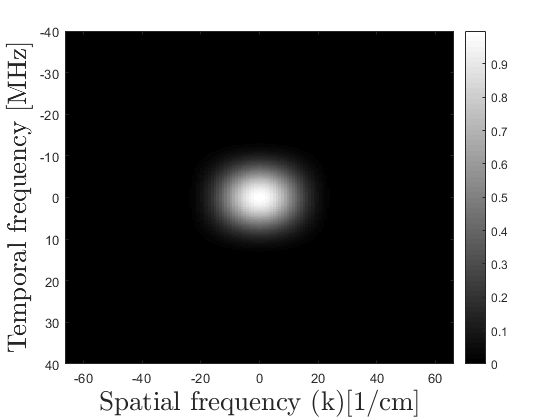}};
    \draw[fill=white] (8.0,-3.9) circle (0.28cm) node[text=black] {c};
     \draw[fill=white] (10,-3.2)  node[text=black] {\large$M_k.N(f)$};
    
     \node[inner sep=0pt] (fftfilt) at (0,-5.5)
      {\includegraphics[width=.35\textwidth]{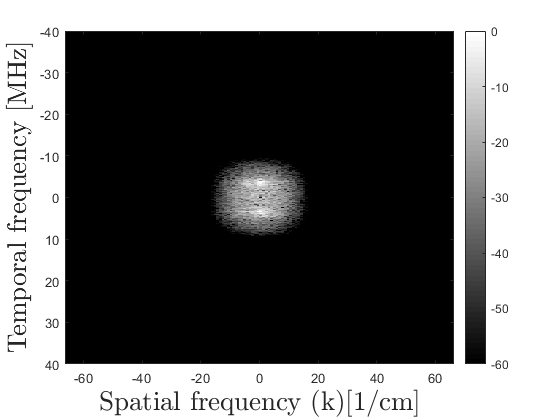}};
      \draw[fill=white] (-2,-3.9) circle (0.28cm) node[text=black] {d};
       \draw[fill=white] (0,-3.2)  node[text=black] {\large$S'_k(f)$};
        \draw[->,line width=0.8mm] (filter.west) -- (fftfilt.east)
          node[midway,fill=white] {\large S*M};
    
    \node[inner sep=0pt] (filt) at (5,-10.7)
         {\includegraphics[width=.35\textwidth]{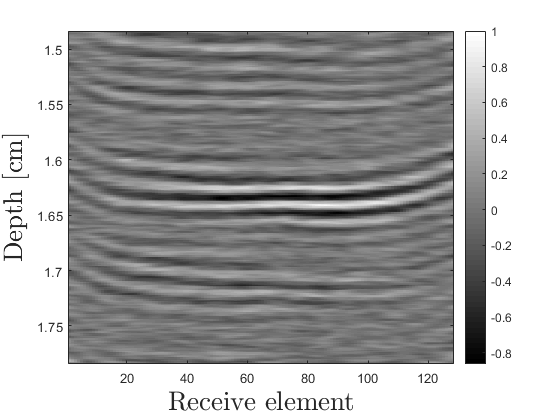}};
         \draw[fill=white] (3.0,-9) circle (0.28cm) node[text=black] {e};
          \draw[fill=white] (5,-8.4)  node[text=black] {\large$s'_i(t)$};
          \draw[->, line width=2mm] (5,-7) -- (5,-8);
            \draw[fill=white] (5.6,-7.4)  node[text=black] {\large iFFT};
    
\end{tikzpicture}
    \caption{aperture domain filtering: (a) transmit beamformed signal with a focal point at speckle point at z=1.63cm. beamforming was performed by all the 128 elements. (b) 2D Fourier transform of the transmit beamformed RF signals. (c) Low-pass filter along aperture and temporal frequency domains. (d) Filtered RF in Fourier domain (e) The final result of filtration with the improved and de-cluttered transmit beamformed data} \label{fig:sub2}
\end{figure}
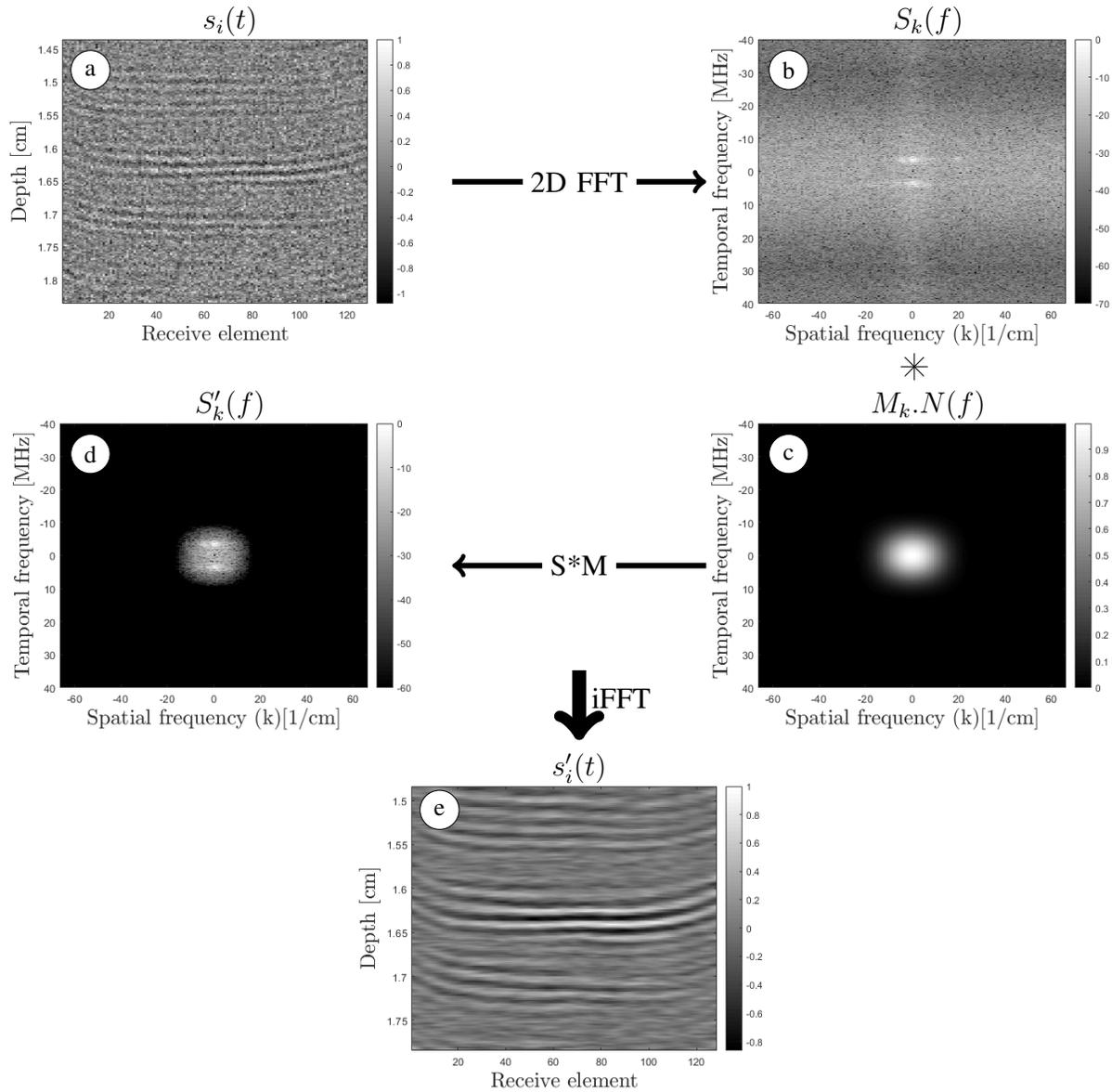

In this research, a 2D Fourier Transform (FT) in the aperture and temporal domain was applied to the 2D TBF RF data, $s_i(t)$. This yielded a Fourier spectrum data (Fig. \ref{fig:sub2}(b)):
\begin{equation} \label{eq 8}
s_{i}(t)  \quad    \quad - 2DFT \rightarrow \quad  \quad  S_{{k}}(f)
\end{equation}
In this formulation, \(k\) represents the aperture frequency corresponding to the aperture dimension, and $f$ is the temporal frequency. $S_k(f)$ is the 2D Fourier transform of $s_i(t)$. It should be noted that after transmit beamforming, and even in presence of phase aberration, the wavefront should exhibit a somewhat horizontal pattern and with only minor distortions. In the spectrum illustrated in Fig. \ref{fig:sub2}(b), horizontal patterns of the aligned wavefronts correspond to the low frequency components along the aperture dimension, such as the bright points around the $k=0$ $cm^{-1}$ axis. Off-axis  clutters (oblique lines the  in Fig. \ref{fig:sub1}(a)) correspond to higher frequency components away from $k= 0$ $cm^{-1}$ axis. Similarly, noise that lacks the horizontal pattern (Fig. \ref{fig:sub2}(a)) had higher frequency components across the Fourier spectrum (Fig. \ref{fig:sub2}(b)). Hence wavefront and noise/clutter can be differentiated by the differences in their aperture frequencies. By applying a low-pass filter in the aperture frequency direction $M_k$, clutter and noise will be reduced significantly. In addition, we found that applying a low-pass filter along the temporal frequency dimension, $N(f)$, could also improve the quality of TBF signals by reducing the noise in the RF signals. Therefore, we constructed a two-dimensional oval-shaped filter (shown in Fig. \ref{fig:sub2}(c)) by multiplying the temporal filter $N(f)$ with the aperture filter $M_{k}$. The filtered spectrum is: 

\begin{equation} \label{eq 9}
 S'_{ {k}}(f) =  S_{ {k}}(f) . M_{k}.N(f). 
\end{equation}

 $M_k$ and $N(f)$ were constructed from Blackman filters to guarantee a gradual cutoff in both the aperture and the temporal frequency directions.
An example of the filtered beamformed spectrum is shown in Fig. \ref{fig:sub2}(d). Finally the filtered TBF signal spectrum \(S'_{ {k}}(f) \) would undergo an inverse Fourier transform along both the temporal and aperture frequency  dimensions. As depicted in the Fig. \ref{fig:sub2}(e) upon filtering the noise was reduced significantly in the filtered TBF signals.  

\begin{figure}
\centering
   \includegraphics[width=.48\textwidth]{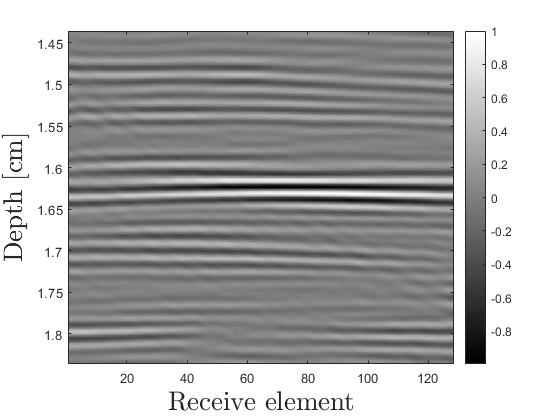}
   \caption{ Transmit beamformed plots after filtration. The off-axis scattering is significalty reduced}
   \label{fig:sub3}
\end{figure}

To demonstrate the effect of the filter on removing the off-axis clutter signals, we applied the filter to the TBF signals in Fig. \ref{fig:sub1}(b). The oblique lines in Fig \ref{fig:sub1}(a) were artifacts generated by the hyper-echoic region that was in the vicinity of the focal point (Fig. \ref{fig:sub1}(b)). It can be seen in Fig. \ref{fig:sub3} that upon filtration the unwanted oblique lines were significantly reduced. It was shown in Figs. \ref{fig:sub2} and \ref{fig:sub3} that the oval filter was capable of removing the off-axis scattering and noise artifacts in the TBF RF data since they lack a horizontal pattern. 
In summary, the oval filters has successfully preserved the shape of the aberrated wavefront while significantly reducing both the noise and clutter in RF signals. NCC will be applied to the filtered TBF RF data that has an improved SNR and SCR (\ref{fig:sub2}(e)) to estimate the phase aberration in the following results section.

\section{Results}
\subsection{Simulation Study }
Filed II ultrasound simulation \cite{JENSEN2006} software was utilized to generate synthetic aperture RF signals. A homogeneous medium with a speed of sound of 1540 \( m/s\) was simulated. Transmit and receive events were simulated with a 128-element transducer having a pitch of 0.15 \(  mm\). The simulation and image reconstruction was performed in Matlab (The MathWorks Inc., Natick, Ma). To mimic a phase aberration scenario, the phase aberration delay profile (truth delay) was created by following the recommendation reported by Dahl et al. \cite{Dahl2005} by convolving a set of Gaussian random numbers with a Gaussian function. Phase aberration was applied to both the transmit and receive process of the STA to simulate the two-way phase aberration in ultrasound propagation in tissues. In the simulations, this ground truth of the phase aberration profile was compared with that estimated from the TBF RF signals. To estimate the aberration delay, NCC algorithm was used, as described by Eq. \ref{eq1} to \ref{eq 5} in Appendix A. To generate noisy data, the STA RF signals were degraded by applying a random noise to individual channels of data to create $RF_{SNR}$ of -10dB. Noise was filtered to match the bandwidth of the ultrasound signals. 

\begin{figure}
\centering
   \includegraphics[width=.48\textwidth]{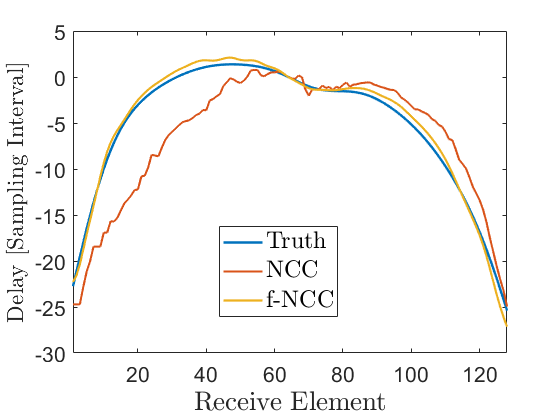}
   \caption{A comparison between the truth, NCC and f-NCC delay estimation in absence of noise. In the NCC method threshold was equal to $\lambda_0/4$. Focal point was at speckle area at 1.63cm depth. Focal point was in vicinity of a hyper-echoic region. }
   \label{fig:sub4}
\end{figure}

In the simulation study, in the absence of noise, the filtered-NCC method was evaluated and compared to NCC as illustrated Fig. \ref{fig:sub4}. In the NCC method a threshold value ($d_{threshold}$) of $\lambda_0/4$ was used. This threshold value is small enough to avoid unwanted large differences between the adjacent error values. Fig. \ref{fig:sub4} shows that the results from NCC (red line) deviates strongly from the ground truth (blue line) for the elements with an index smaller than 80 (on the left side of the delay profile) while the results of f-NCC (the yellow line) agrees well with the ground truth. It is noticed that the off-axis clutter signals (the oblique lines) in Fig. \ref{fig:sub1}(b) intersect with the wavefront of interest (around the focal depth of 1.63cm) mainly at elements with index smaller than 80. Therefore, we believe that the clutter signals caused the deviation of the results (Fig. \ref{fig:sub4} ) (red line) from the ground truth (blue line). As illustrated in Fig. \ref{fig:sub3}, the Oval filter in F-NCC has eliminated the oblique lines. Therefore, f-NCC has estimated the delay profile accurately.

\begin{figure}
\centering
   \includegraphics[width=.48\textwidth]{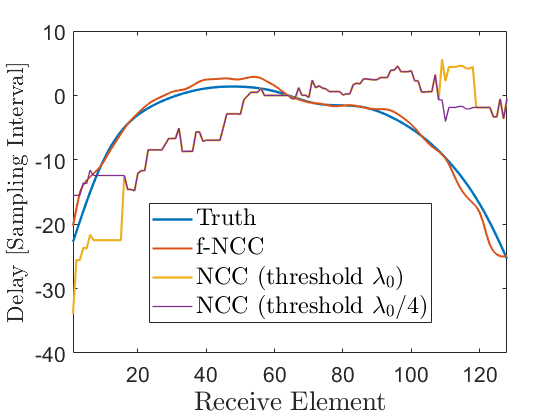}
   \caption{A comparison between the truth, NCC, and f-NCC delay estimation. Independent noise was added to the ultrasound signal to achieve a ${SNR}$ of -10 dB. NCC method was coupled with thresholds values of $\lambda_0$ and $\lambda_0/4$ for adjacent element values. Focal point was on a speckle region at the depth of 1.63cm }
   \label{fig:sub5}
\end{figure}

Later on, the iterative method was applied to the noisy RF signals when focal point was located at the same depth ($z = 1.63 cm$) over an speckle area. Fig. \ref{fig:sub5} shows that NCC estimation of phase aberration delays are different from the truth as noise was introduced to the RF signal. Ideally, the difference between the adjacent elements in delay error  should be small and the delay profile should have a smooth pattern. However, in the presence of noise, NCC fails and can result in a large difference between the adjacent elements in  delay errors. To improve NCC, threshold values of $\lambda_0$ and $\lambda_0/4$ were applied to the adjacent element delay errors as explained in Eq.\ref{eq 19} in appendix A. Fig. \ref{fig:sub5} illustrates that the NCC along with two different values of threshold could not estimate the error delays accurately. As shown in Fig. \ref{fig:sub2}, oval filter has removed the superposed noise from the TBF RF data. Consequently, f-NCC (shown by red line in Fig. \ref{fig:sub5}) were able to estimate the aberration delay profile accurately. \textbf{In Fig. \ref{fig:sub5}, a comparison between truth, NCC with two threshold values and f-NCC delays profiles proves the success of this filtering method to resolve the limitation of NCC. } 

The sizes of the oval filter, especially the cutoff frequency in the aperture direction, are important factors. Various window sizes for the aperture and temporal filter were tested.  Filters with excessively small cutoff frequencies would over-flatten and alter the overall shape of the wavefront. This is caused by the removal of the desirable frequencies in the Fourier domains around $k=0$ $cm^{-1}$ and $f=0 $ $MHz$. On the other hand, For large cutoff windows, filter would allow passage of unwanted frequencies and therefore the noise and off-axis scattering reduction would not be successful. In this case, estimated delay profile would be distorted. A decision for the cutoff frequency was a trade-off between the smoothness of the profile and preserving the shape of the wavefront. In this study cut-off frequencies of \(18 cm^{-1}\) and $12 MHz$ for oval filters were used. 

Due to the periodic nature of Fourier transform, the edge effect was present in the delay profile estimation, as shown in Fig. \ref{fig:sub11} in appendix B. To correct the distortions in the edges of the delay profile an iterative padding technique was used. See details in appendix B.

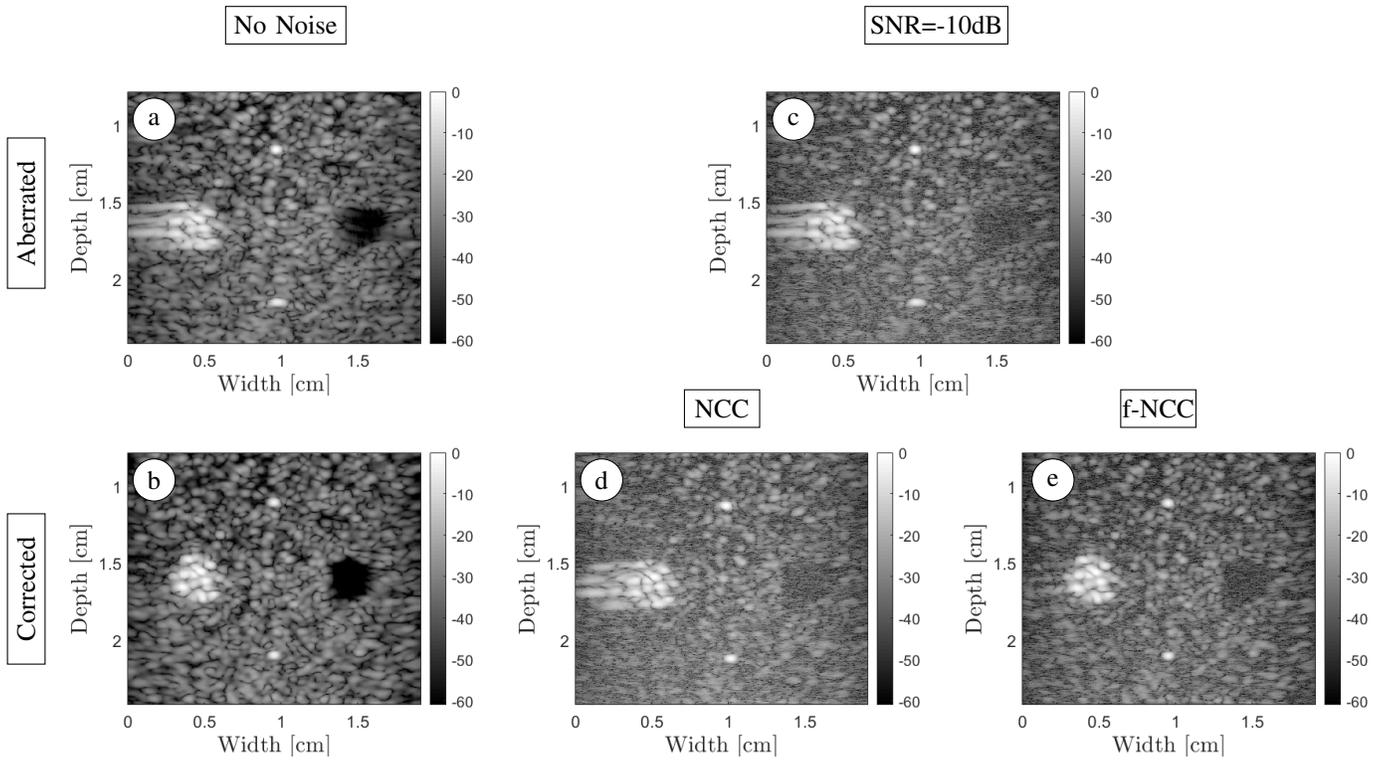
\begin{figure}
\centering
\begin{tikzpicture}

\node[inner sep=0pt] (org) at (0,0)
    {\includegraphics[width=.32\textwidth]{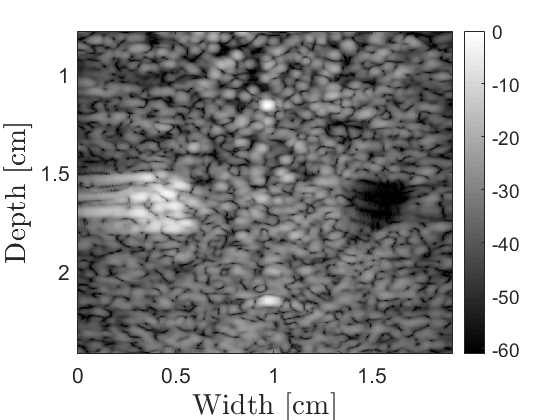}};
    \draw [fill=white](-0.8,2.5) rectangle (0.8,3) node[pos=.5] {No Noise};
   \draw [fill=white](-3.2,-0.75) rectangle (-3.7,1.25) node[pos=.5, rotate=90 ] {Aberrated};
   \draw[fill=white] (-1.75,1.5) circle (0.28cm) node[text=black] {a};
  
       \node[inner sep=0pt] (filter) at (0,-4.8)
    {\includegraphics[width=.32\textwidth]{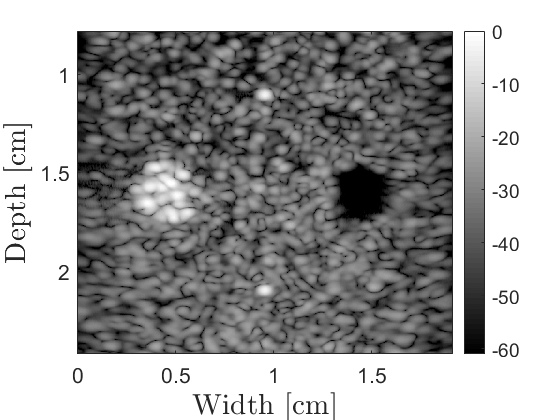}};
      \draw [fill=white](-3.2,-5.75) rectangle (-3.7,-3.75) node[pos=.5, rotate=90 ] {Corrected};
    \draw[fill=white] (-1.75,-3.3) circle (0.28cm) node[text=black] {b};
  
 \node[inner sep=0pt] (fft) at (8.5,0)
    {\includegraphics[width=.32\textwidth]{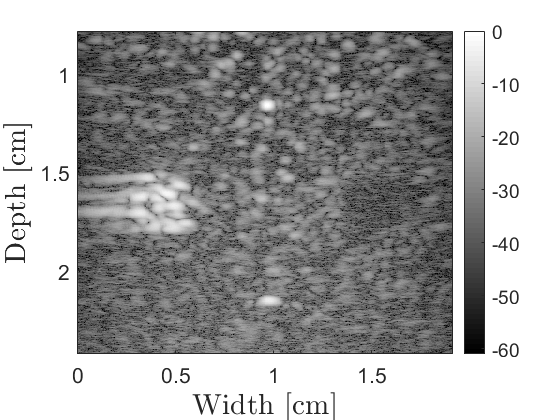}};
   \draw [fill=white](7.7,2.5) rectangle (9.6,3) node[pos=.5] {SNR=-10dB };
  \draw[fill=white] (6.75,1.5) circle (0.28cm) node[text=black] {c};

     \node[inner sep=0pt] (fftfilt) at (5.95,-4.8)
      {\includegraphics[width=.32\textwidth]{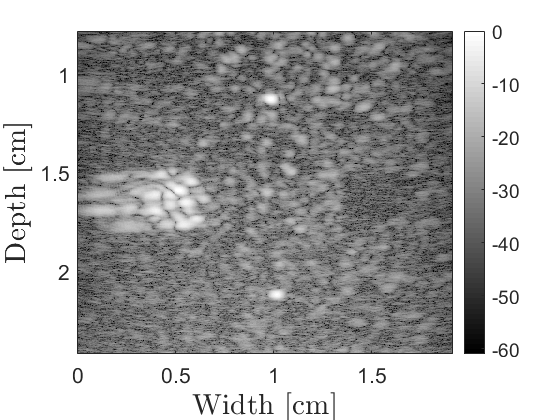}};
\draw[fill=white] (4.2,-3.3) circle (0.28cm) node[text=black] {d};
    \draw  (5.3,-2.6) rectangle (6.3,-2.1) node[pos=.5 ] {NCC};
    
    \node[inner sep=0pt] (filt) at (11.9,-4.8)
         {\includegraphics[width=.32\textwidth]{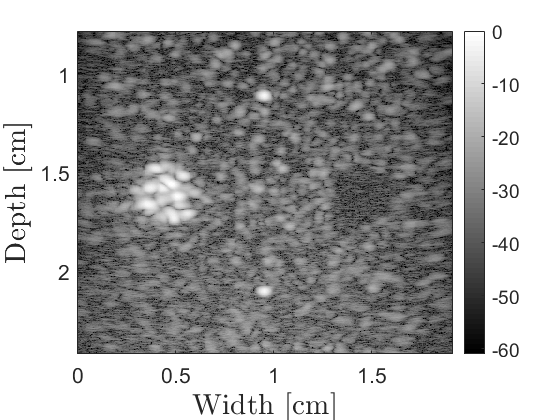}};
      \draw[fill=white] (10.2,-3.3) circle (0.28cm) node[text=black] {e};
   \draw  (11.1,-2.6) rectangle (12.1,-2.1) node[pos=.5] {f-NCC};
    
\end{tikzpicture}
   \caption{ STA images reconstructed from : (a) the noiseless aberrated data, (b) the NCC-corrected noiseless data, (c) noisy RF signals with aberration, (d) the NCC-corrected noisy data, and (e) the f-NCC-corrected noisy data. Focal point was located at speckle point at z=1.63 cm}
   \label{fig:sub7}
\end{figure}

Next, Fig.  \ref{fig:sub6} compares the performance of NCC and f-NCC algorithms in TBF data of RF signals without noise and when $RF_{SNR}=-10dB$.  Fig.  \ref{fig:sub7} shows their corresponding STA images. Fig.  \ref{fig:sub6}(a) shows the noiseless, aberrated TBF RF data.  Fig. \ref{fig:sub6}(b) illustrates that NCC has improved the alignment of wavefront by correcting the phase aberration. Even though in Fig. \ref{fig:sub4}, NCC delay profile is different from the truth, especially on the left side of the graph, there is a general agreement in the rest of the graph. Consequently, NCC delay error compensation has resulted in an improved wavefront in  Fig. \ref{fig:sub6}(b). This observation is compatible with the reconstructed images in Fig. \ref{fig:sub7}(a) and b. In Fig. \ref{fig:sub7}(a) the smearing of the hyper-echoic lesion, fill-ins into the hypo-echoic area, and the out-of-focus point reflector at the depth of 2.5cm are  indications of phase aberration. Even though a misalignment can be seen in Fig.  \ref{fig:sub6}(b), NCC has reduced the phase aberration effect significantly for a noiseless RF in Fig. \ref{fig:sub7}(b). In this image the lesions and point reflectors are sharper and has better delineation. 
 
 Later on, NCC was investigated for a noisy RF ($RF_{SNR}=-10dB$). The TBF data from the noisy and aberrated data was plotted in Fig. \ref{fig:sub6}(c). The corrected RF data were presented in  Fig. \ref{fig:sub6}(d), showing the poor performance of NCC for noisy RF data. As illustrated in Fig. \ref{fig:sub5}, NCC is not capable of estimating a correct phase aberration delay for noisy RF (yellow and purple lines). The image reconstructed from the noisy and aberrated data in Fig. \ref{fig:sub6}(c) was shown in Fig. \ref{fig:sub7}(c). There is obvious evidences of phase aberration and noise in this image. The image reconstructed from the noisy data that had been corrected with NCC  is shown in Fig. \ref{fig:sub7}(d). In this image, the effect of phase aberration is still present. Even though the focal points are sharper,  the hyper-echoic lesion is smeared and the contrast is not improved. 
 
Fig. \ref{fig:sub6}(e) shows the TBF data after filtration with the designed oval filter. The noise has been significantly reduced and wavefront is more apparent. The f-NCC delay profile was estimated from this TBF RF data and as illustrated in Fig. \ref{fig:sub5}, it is compatible with the truth profile. A comparison between Fig. \ref{fig:sub6}(d) and Fig. \ref{fig:sub6}(f) indicates the advantage of f-NCC versus NCC methods to flatten the wavefront. After f-NCC delay compensation, the abberated wavefront shown in Fig. \ref{fig:sub6}(e) was successfully aligned as illustrated in Fig. \ref{fig:sub6}(f). It should be noted that the corrected TBF RF data plots was achieved after 2 iterations. The corrected image with f-NCC is plotted in Fig. \ref{fig:sub7}(e).
Fig. \ref{fig:sub7} (c), (d) and (e) shows that f-NCC was successful in correcting the phase aberration effects from the images: 1) image \ref{fig:sub7}(e) has better contrast and delineates the lesions better. 2) Fig. \ref{fig:sub7}(c) and (d) show blurring in speckles whereas this issue is significantly improved after f-NCC correction in image (e). 
Additionally, the effect of phase aberration correction on the images was quantitatively compared by measuring the peak signal-to-noise ratio (PSNR) of the point reflector and the contrast-to-noise-ratio (CNR) values of the hyper-echoic region. As shown in Table. \ref{tab1} values of PSNR and CNR have supported the above claims. After the f-NCC corrections, CNR\(_{hyper}\) and PSNR  have been improved by 24 and 51\%, respectively. 
 
\subsection{Experimental Study}
\subsubsection{Experiment Setup/Method}
\begin{figure}
\centering
\subfloat[][]{\includegraphics[width=.23\textwidth]{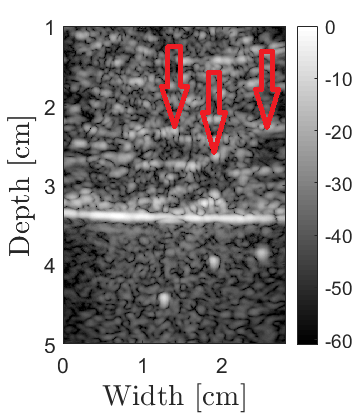}} 
   \subfloat[][]{\includegraphics[width=.24\textwidth]{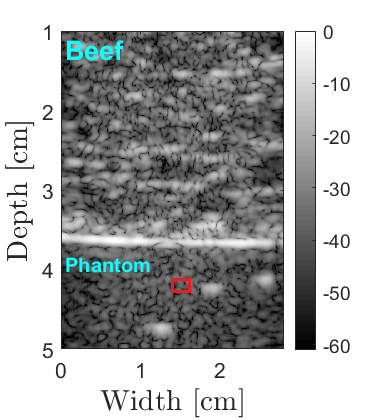}} \\
   \subfloat[][]{\includegraphics[width=.23\textwidth]{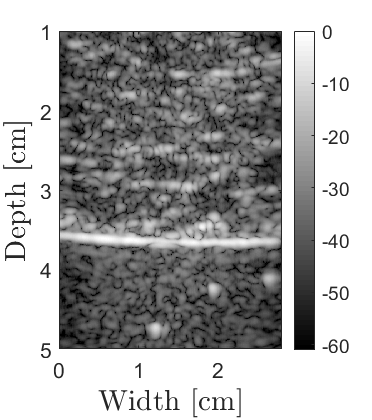}}
   \subfloat[][]{\includegraphics[width=.25\textwidth]{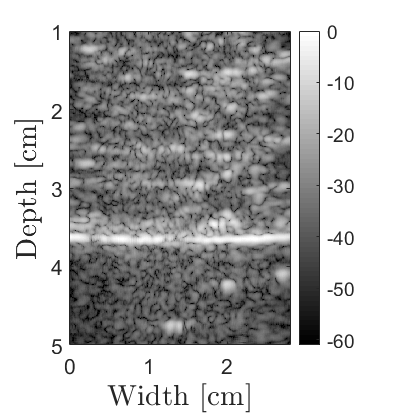}}
   \caption{Experimental  STA  images reconstructed from  RF signals with (a) a correct speed of 1540 m/s and (b) a wrong speed of 1650 m/s. STA images reconstructed from the  RF signals corrected with (c) f-NCC and (d) NCC methods. A beef top layer was added to create a noisy RF data. The focal point to estimate the phase aberration (signified by the red square) was at 4.1 cm.}
   \label{fig:sub9}
\end{figure}

The proposed method was investigated on a tissue-mimicking phantom, CIRS calibrated phantom (Multi-purpose multi-tissue ultrasound phantom Model 040 GSE) with nominal sound-speed of 1540 \(\pm\) 20 m/s. A segment with a width of 2.95cm was scanned. To create a noisy RF data, CIRS phantom was topped with a 3.5-cm thick slab of beef, which had a speed of sound of about 1540-1560 m/s \cite{Goss1978}. This calculated speed is similar to the value for the calibrated phantom. Therefore it was reasonable to assume the phase aberration caused by the beef is minor. The area of study within the CIRS phantom contained 3 point reflectors(Fig. \ref{fig:sub9}(a)). Additionally, three hypo-echoic regions,  which are signified with the red arrows, are present within the beef area. Synthetic aperture raw RF data was acquired with Verosonics Vantage scanner and a P6-4 phased array probe with a central frequency of 3.47 MHz. Sampling frequency was four times the central frequency. Later on, the RF signals were up-sampled to 55.5 MHz in software. The probe was fixated by an articulated arm and placed directly on top of the phantom. The speckle focal point was located in the CIRS phantom (under the beef), at the imaging depth of 4.1 cm. The ground truth of the sound speed \(c_t\) was assumed to be 1540 m/s. 

Unlike the simulation study, it is challenging to know the true phase aberration of an experimental tissue-mimicking phantom. \textit{Therefore, in the experimental study the proposed phase aberration estimation methodology was studied under a speed error scenario, in which a wrong sound speed  of 1640 m/s was intentionally used to reconstruct the STA images.} In the formulation, we assumed that the phase aberration other than that caused by the speed error can be neglected. The advantage of a speed error scenario is that the phase aberration delay profile can be formulated to quantitatively compare NCC and f-NCC methods. Additionally, in the speed error scenario phase aberration is a function of depth, as shown in Eq. \ref{subeq2}. This makes an ideal case to investigate the f-NCC method for STA data. Since NCC can only estimate the relative shifting between elements, the phase aberration of the $i_{th}$ element was derived relative to (by subtracting) that of the element right above the focal point (point C in Fig. \ref{fig:sub12} of the appendix). The true delay error is formulated as below:

\begin{subequations}
\begin{align}
    \delta t_{i}=  \delta t_c\cdot (1+cos\theta_i) \label{subeq1}\\
    \delta t_c =  l_{i}*(1-cos \theta_i) *  [\frac{1}{c_r}-\frac{1}{c_t}] \label{subeq2}
\end{align}
\end{subequations}

where \(i\) represents the receive element index, \(\delta t_{i}\) is the phase aberration error of the $i_{th}$ element relative to that of the element right above the focal point, \(l_{_i}\) is the travel distance between the focal point and receive element \(i\), and \( \theta_i\) is the angel between the travelled path to the receive element \(i\) and the norm to the probe, as shown in Fig. \ref{fig:sub12} of the appendix. \(c_t\) represents the true average speed of sound and \(c_r\) is the reconstruction sound speed. $\delta t_c$ represents the part of the $i_{th}$ element phase aberration (relative to the element right above the focal point) caused directly by the sound speed error.

The \(cos \theta_i\) term in Eq. \ref{subeq1} is an unexpected term. The \(cos \theta_i\) term is due to the vertical shift of the focal point caused by the speed error in the image reconstruction. Considering that $\theta_i$ usually small, the $cos \theta_i$ term is close to 1. The derivation of the formulation will be explained in appendix \ref{speed error}. The above formula was validated with Field II simulations and there was a good agreement between the theory and simulations.

Oval filter with the window size of \(11 cm^{-1}\) in aperture frequency domain and \(17 MHz\) in temporal frequency domain was found as the optimum windows in this experimental study. Ultimately, the experimental RF was corrected over transmit and receive.

\begin{figure}
\centering
   \includegraphics[width=.45\textwidth]{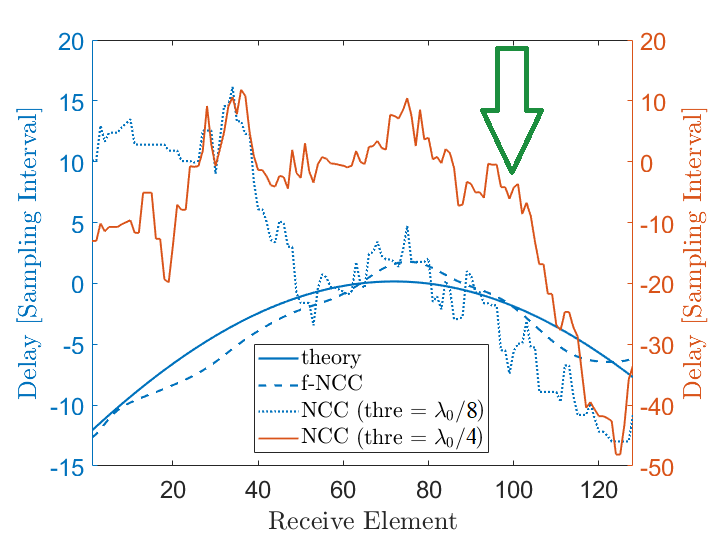}
   \caption{ A comparison between the theory and the estimated phase aberration. NCC coupled with thresholds of $\lambda_0/4$ and $\lambda_0/8$ and f-NCC were included. Theory delay profile was calculated using Eq. \ref{subeq2} with respect to focal point at 4.1 cm. }
   \label{fig:sub8}
\end{figure}

Fig. \ref{fig:sub8} shows the importance of the filtration to achieve a proper delay profile estimation. Signals were originated from a speckle area at the depth of 4.1cm. In this Fig. $c_t$ was assumed to be 1540 m/s. The NCC method was coupled with two values of threshold. The delay profile corresponding to the threshold value of $\lambda_0/4$ exhibits undesirable jumps in the profile. Nevertheless, it still shows a somewhat similar trend to the true  delay errors in the second half of the delay profile (orange line) indicated by the green arrow. This suggests the possibility that a smaller threshold value could fix the problem. However NCC coupled with a smaller threshold ($\lambda_0/8$) shows a substantial difference from the theory. This plots proves the success of f-NCC estimation when compared to the theory profile (Eq. \ref{subeq2}) with 1540 m/s as the true speed. The small differences between the theory and f-NCC estimation might be due to the existence of the spatial speed heterogeneity within the beef layer. These sound speed variations contribute to the overall phase aberration delay. As indicated earlier, Eq. \ref{subeq2} only formulates the phase aberration due to the average speed error and not the spatial speed variation within the beef. To verify this claim, we collected RF data of the CIRS phantom in the absence of beef layer and estimated the phase aberration for a speed error scenario. The f-NCC estimated delay profile agreed well with the theory.
 
 First, a phase aberration-free image (Fig. \ref{fig:sub9}(a)) was reconstructed with the correct speed of 1540 m/s. Second, the aberrated image with sound speed error is illustrated in Fig. \ref{fig:sub9}(b). A comparison between Fig. \ref{fig:sub9}(a) and b shows that phase aberration degradation is present in the STA image with speed error as: 1) the point reflectors within the phantom are unfocused and 2) the three dark areas within the beef at around depth of 3cm have fill-ins. The dark areas are shown by red arrows on Fig. \ref{fig:sub9}(a). Images corrected by f-NCC and NCC With the focal point at 4.1cm deep are shown in Fig. \ref{fig:sub9}(c) and (d), respectively. 

The comparison between these images of Fig. \ref{fig:sub9} are in agreement with the delay profiles in Fig. \ref{fig:sub8}. F-NCC was more effective than NCC in improving image quality because in Fig. \ref{fig:sub9}: 1)image (c) exhibits a more improved spatial resolution, specially for phantom point reflectors at approximately 4 cm and 5 cm; 2)in image (c), beef structural features at around 3cm (shown by red arrows in image (a) has better contrast and delineation; 3)the interface between beef and the CIRS phantom in image (c) is sharper while it has a broken structure in Fig. \ref{fig:sub9} (d). 4)speckles in image c has a more uniform and sharper structure than Fig. \ref{fig:sub9}(d), where speckles have a smaller scale and broken pattern. This conclusion has been confirmed by image quality metrics. F-NCC corrections has improved the PSNR by 24\% whereas this improvement is 5\% after NCC correction.

It should be pointed out that the phase aberration induced by sound speed error depends on the position of the focal point or pixel of interest. In the above correction, we chose the focal point to be a speckle at $x=1.1cm$  and $z= 4.1 cm$. Therefore, among the three point reflectors the sharpness of the nearby point reflector at position (2, 4.1) cm was improved the most. Alternatively, if we find the phase aberration by focusing on speckles around other point reflectors, the sharpness of the corresponding reflectors will be improved the most. In practice, it is more efficient to correct the phase aberration caused by the speed error by adjusting the average speed in the image reconstruction, rather than by finding the phase aberration point by point. Here we use the speed error model to demonstrate the ability of the proposed method to correct phase aberration pixel by pixel in the image. 

\section{Discussion}
 In the simulation and experimental study we proved the feasibility of f-NCC to estimate and correct the phase aberration error. A potential application of this filtering technique is to de-clutter the TBF RF data in B-mode imaging, prior to receive beamforming. Consequently, phase aberration in B-mode can be estimated after TBF RF data is de-cluttered. It should be noted that, f-NCC can determine phase aberration most accurately around the focal point of the transmission beam. For speckles that is away from the focal point, the performance of NCC will be compromised. This is due to the decrease of the correlation between the TBF RF signals from different elements, which is caused by the large beam width outside the focal zone \cite{Mallart1991}. In contrast, STA imaging will not suffer this limitation because every pixel in STA image is optimally focused. Nevertheless, the designed filter has the potential to be used in conjunction with other algorithms that require noise/clutter reduction as a prior step in B-mode imaging.
 
Another potential application of  Eq. \ref{subeq2} that is under study is to develop an  algorithm to estimate the average speed of a medium. With f-NCC proven to be an effective estimator of the delay profile $t_i$, this equation can be rearranged to solve for the scalar value $c_t$. This problem can easily be solved and with a low computational load.  

Another potential application of the f-NCC algorithm is to develop an  algorithm to map the  local speed of a medium. This can be done based on the fact that, at any speckle point, the phase aberration corresponding to the array elements is the result of the travel time change that the signal has experienced along the travelled path. This path is the line connecting the pixel/speckle and the array elements. The final delay error at every speckle can be represented as the integration of the sound speed error over the path \cite{Jakovljevic2018}. Once we estimate the phase aberration delay of many speckles in an ultrasound image, we can write down a linear equation set that express the phase aberration in terms of the local speed error. After that, the standard methods in solving inverse problems can be applied to obtain the local speed. We are developing a method along this approach to map the local speed of tissues with STA.

\section{Conclusion}
 We demonstrated the efficacy and robustness of an iterative f-NCC method in STA to estimate the phase aberration at speckle area from single-frame noisy RF signals in both simulation and experimental study. The 2D filtration of the TBF data was performed in the aperture and temporal domains to partially remove off-axis signals and noise. An iterative zero-padding technique was developed to correct the edge effect caused by filtering.
A new formula of the phase aberration  profile was derived and validated when a wrong average sound speed was used in the image reconstruction.  The proposed method can improve the image quality of STA images with phase aberration. Alternative potential applications of the method include average speed estimation and local speed mapping with STA.

 \appendices
 \section{Using Normalized-Cross-correlation to estimate the residual phase aberration $r_{{i}}$}
  \label{NCC}
 Normalized-Cross-correlation is a window matching function that quantifies the relative shift between two similar signals \(s_{i}\) \& \(s_{i+p}\) \cite{ODonnell1988a}:

\begin{equation} \label{eq1}
NCC(\tau)=\frac{\sum_{t=w_{1}}^{w_{2}} s_{{i}}(t) s_{i+p}(t+\tau)}{ \sqrt {\sum_{t=w_{1}}^{w_{2}} s_{{i}}^2 (t)  \sum_{t=t_{1}}^{t_{2}} s_{i+p}{^2(t+\tau)}} }
\end{equation}

Here \(w_{2}\) \&  \(w_{1}\) represent the starting and ending point of the window in the signals \(s_{i}\) \& \(s_{i+p}\), over which cross correlation is measured. Additionally, \(\tau\) is the time shift between the two signals. Parameter \(p\) is the receive element lag between the cross-correlated signals (e.g., $p=1$ indicates the neighboring elements). The shift \(\hat{d}_{i,i+p}\), between the receive elements with lag $p$ is at the peak of the cross-correlation function:

 \begin{equation} \label{eq2}
\hat{d}_{i, i+p}=\underset{\tau}{\arg \max } (NCC(\tau))
\end{equation}

In this implementation, signals were up-sampled to 80 MHz and \textit{sub-sampling accuracy in NCC algorithm was achieved with cosine curve fitting.} NCC was performed between a pair of receive elements with lag 1 (adjacent elements) \& lag 2. 

 In theory, we have 

\begin{equation} \label{eq3}
\hat{d}_{{i+1}, {i}}=r_{{i+1}}-r_{{i}},
\end{equation}

where  \(r_{{i}}\) and \(r_{i+1}\) are the residual delay errors corresponding to the \(i_{th}\) and \((i+1)_{th}\)receive elements . 

In this paper, we assume that the residual phase aberration profile \(r_{{i}}\) is a smooth function of $i$. Therefore, we expect $\hat{d}_{{i}}$ to be small value compared to the wavelength according to Eq. \ref{eq3}.   However, the noise in the RF signals would cause a drastic increase in some adjacent delay error calculations and jeopardize an ideally smooth delay profile.  A straightforward solution was to define a threshold for  $\hat{d}$:  

\begin{equation} \label{eq 19}
\hat{d_i}=0 \quad  if    \quad \hat{d_i}> d_{threshold}.
\end{equation}

Eq. \ref{eq3} can be written into a matrix form as below:

\[ \quad \quad  \quad    \quad   \quad   M \quad   \quad  \quad \quad  \quad   \quad  \quad  \quad   r \quad\quad \quad   \quad  \quad  \quad  d\]
\begin{equation} \label{eq4}
\left[\begin{array}{cccccc}{1} & {-1} & {0} & {0} & {\cdots} & {0} \\ {0} & {1} & {-1} & {0} & {\cdots} & {0} \\ {0} & {0} & {1} & {-1} & {\cdots} & {0} \\ {\vdots} & {} & {} & {} & {\vdots} \\ {0} & {0} & {\cdots} & {0} & {1} & {-1}\end{array}\right]\left[\begin{array}{c}{r_{1}} \\ {r_{2}} \\ {r_{3}} \\ {\vdots} \\ {r_{N-1}}\end{array}\right]=\left[\begin{array}{c}{d_{1,2}} \\ {d_{2,3}} \\ {d_{3,4}} \\ {\vdots} \\ {d_{N-1, N}}\end{array}\right]
\end{equation}

The true arrival time error was estimated by taking the pseudo-inverse of the above equation:

\begin{equation} \label{eq 5}
\hat{r}=\left(M^{T} M\right)^{-1} M^{T} d
\end{equation}
The estimation can be improved if M includes both lag 1 and lag 2 shift. That can be achieved when the corresponding matrices of \(M\) and \(d\) for both lags are stacked. 

In summary, Eq. \ref{eq1}  was first used to yield the shift between a pair of receive elements \(d_{i},_{i+p}\) with the desired lag \(p\). The threshold value was applied according to equation \ref{eq 19}. Then the delay error for each receive element can be computed by solving Eq. \ref{eq 5}. After that, the estimated delay was compensated to the raw RF data for both the transmit and receive process to approximately correct the phase aberration. Lastly, the above process was iterated several times for further improvement. This technique takes approximately 2-3 iterations to reach convergence and proper correction of the phase aberration error. 

\section{Padding}
\begin{figure}
\centering
   \includegraphics[width=.45\textwidth]{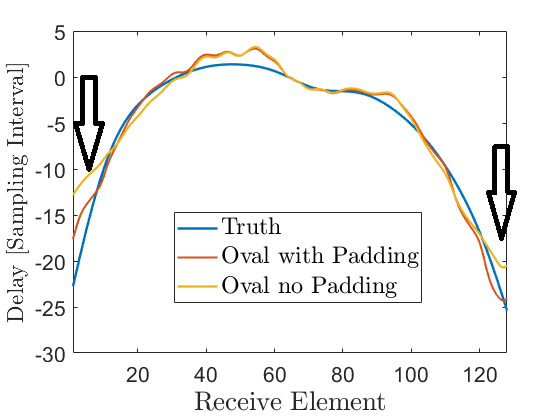}
   \caption{A comparison between the truth and estimated delay profiles to show the effectiveness of padding to correct the edge effect. In this graph blue line represents the Truth delay profile. Red and yellow lines compare the corrected profiles with oval filter with and without padding.}
   \label{fig:sub11}
\end{figure}

In the estimation of the delay profile, a much larger error  can be found at the edge in f-NCC method, as shown by the black arrows at the elements 1-10 and 118-128 in Fig. \ref{fig:sub11}. This is caused by the periodic assumption of signals when applying Fourier transform. Based on the periodic assumption, signal from channel 128 will be wrapped round to be adjacent to the signal from channel 1. Since the signals from these two channels are quite different, there is a discontinuity in the 2D TBF RF signals. After going through the low-pass filters, the locations with discontinuity will result in larger distortion compared to other parts. To accommodate such problems, we used an approach borrowed from the zero-padding technique in signal processing field \cite{kak2002}. An iterative padding scheme was proposed. In our implementation of the filtering process, the 2D TBF RF signals were stored in a matrix $S_i(t)$, each column represented the signal from one channel. In the first iteration, TBF signal matrix was padded by replicating several times, the signal of the first channel on the left side of the matrix and the signal of the $128_{th}$ channel on the right side of the matrix. Then this expanded matrix was filtered as described in the Section II A. After that, f-NCC was applied to calculate the delay profile. In further iterations, signals were extrapolated on both sides with delays that have the same slope on the delay profile as these of the first and last 10 elements. Fig. \ref{fig:sub11} illustrates that this iterative padding technique has minimized the edge effect.

\section{Formulation of Delay Profile under a Speed Error Scenario}
  \label{speed error}

To validate the f-NCC methodology, the true phase aberration delay was derived for a scenario when image was reconstructed with a wrong average speed. The speed error will result in travel time error in two ways: first, it will affect the travel time by the wrong speed directly. Second, it will shift the target vertically in the image, which in turn will cause an additional travel time error. Initially, for the focal point A (Fig. \ref{fig:sub12}), the travel time relative to the element right above the focal point was calculated. Later on, $l_{AC}$ was subtracted so element right above the focal pixel has no delay :

\begin{equation} \label{eq 21}
t_{c} = \frac{l_{AD}-l_{AC}}{c}= \frac{ l_i}{c}* (1-cos\theta_i)
\end{equation}

For the simplicity in this formulation, the path with respect to point $A$ was shown by $ l_i$. In the above equation, $\theta_i $ is the angel between travel path with respect to receive element $i$ and norm to the path. The goal of this calculation was to find the delay error time corresponding to element $i$, located at point $D$. Under a speed error condition, the delay error is:

\begin{equation} \label{eq 12}
 \delta t_{c} = l_i * (1-cos\theta_i) *  [\frac{1}{c_r} -\frac{1}{c_t}]
\end{equation}
$c_r$ represents the reconstruction speed and $c_t$ is the true speed.
Additionally, in presence of speed error the focal point A is shifted by $l_{AB}$ or $\delta l$:

\begin{equation} \label{eq13}
\delta l=l_{AB} = (l_i - \frac{l_i}{c_t}.  c_r) * cos \theta_i = l_i . c_r *  [\frac{1}{c_r}-\frac{1}{c_t}] * cos \theta_i
\end{equation}
Time error due to the vertical shift will be:
    
\begin{equation} \label{eq 14}
 \delta t_{d} =  [\frac{l_{DB}-l_{BC}}{c_r}-\frac{l_{AD}-l_{AC}}{c_r}]
\end{equation}
This equation was rearranged as below : 

\begin{equation} \label{eq 15}
 \delta t_{d} =  [\frac{l_{DB}-l_{AD}}{c_r}-\frac{l_{BC}-l_{AC}}{c_r}]
\end{equation}

By substituting for the position shift $\delta l$ this equation was obtained: 

\begin{equation} \label{eq 16}
\begin{aligned}
 \delta t_{d} =  \frac{- \delta l cos(\theta_i)+\delta l}{c_r}=\frac{\delta l [1-cos(\theta_i)]}{c_r}=  
 \\
 \frac{l_i * c_r * cos \theta_i [1-cos(\theta_i)]}{c_r}* [\frac{1}{c_r}-\frac{1}{c_t}]=\delta t_c. cos\theta_i
 \end{aligned}
\end{equation}
To find the total delay error, speed error $\delta t_c$  and position shift $\delta t_d$ delays were added: 

\begin{equation} \label{eq 17}
\delta t_{i}= \delta t_{d} +\delta t_c=  \delta t_c. (1+cos\theta_i)
\end{equation}

 \section*{Acknowledgment}
The authors would like to thank the following funding agencies: Natural Sciences and Engineering Research Council of Canada (NSERC), Canada Foundation for Innovation (CFI), and Ryerson University. We are grateful to Dr. Michael Kolios, Dr. Ying Li, Dr. Ping Gong, and Na Zhao for their valuable discussions and suggestions. We thank Dr. Micheal Kolios for providing access to his research facility in this study.

\bibliographystyle{IEEEtran}
\bibliography{phase.bib}

% Generated by IEEEtran.bst, version: 1.14 (2015/08/26)
\begin{thebibliography}{10}
\providecommand{\url}[1]{#1}
\csname url@samestyle\endcsname
\providecommand{\newblock}{\relax}
\providecommand{\bibinfo}[2]{#2}
\providecommand{\BIBentrySTDinterwordspacing}{\spaceskip=0pt\relax}
\providecommand{\BIBentryALTinterwordstretchfactor}{4}
\providecommand{\BIBentryALTinterwordspacing}{\spaceskip=\fontdimen2\font plus
\BIBentryALTinterwordstretchfactor\fontdimen3\font minus
  \fontdimen4\font\relax}
\providecommand{\BIBforeignlanguage}[2]{{%
\expandafter\ifx\csname l@#1\endcsname\relax
\typeout{** WARNING: IEEEtran.bst: No hyphenation pattern has been}%
\typeout{** loaded for the language `#1'. Using the pattern for}%
\typeout{** the default language instead.}%
\else
\language=\csname l@#1\endcsname
\fi
#2}}
\providecommand{\BIBdecl}{\relax}
\BIBdecl

\bibitem{Tabei2003}
M.~Tabei, T.~Mast, and R.~Waag, ``Simulation of ultrasonic focus aberration and
  correction through human tissue,'' \emph{Journal of the Acoustical Society of
  America}, vol. 113, no.~2, pp. 1166--1176, 2003.

\bibitem{Guenther2009}
D.~Guenther and W.~Walker, ``Generalized cystic resolution: A metric for
  assessing the fundamental limits on beamformer performance,'' \emph{IEEE
  Transactions on Ultrasonics, Ferroelectrics, and Frequency Control}, vol.~56,
  no.~1, pp. 77--90, 2009.

\bibitem{Mann2002}
J.~Mann and W.~Walker, ``A constrained adaptive beamformer for medical
  ultrasound: Initial results,'' vol.~2, 2002, pp. 1807--1810.

\bibitem{Bell2015}
M.~Lediju~Bell, J.~Dahl, and G.~Trahey, ``Resolution and brightness
  characteristics of short-lag spatial coherence (slsc) images,'' \emph{IEEE
  transactions on ultrasonics, ferroelectrics, and frequency control}, vol.~62,
  pp. 1265--1276, July 2015.

\bibitem{Chau2017}
G.~Chau, J.~Dahl, and R.~Lavarello, ``Effects of phase aberration and phase
  aberration correction on the minimum variance beamformer,'' \emph{Ultrasonic
  imaging}, vol.~40, pp. 16--17, July 2017.

\bibitem{ODonnell1988}
M.~O'Donnell and S.~W. Flax, ``{Phase-Aberration Correction Using Signals From
  Point Reflectors and Diffuse Scatterers: Measurements},'' \emph{IEEE
  Transactions on Ultrasonics, Ferroelectrics, and Frequency Control}, vol.~35,
  no.~6, pp. 768--774, 1988.

\bibitem{Krishnan1996}
S.~Krishnan, P.-C. Li, and M.~O'Donnell, ``Adaptive compensation of phase and
  magnitude aberrations,'' \emph{IEEE Transactions on Ultrasonics,
  Ferroelectrics, and Frequency Control}, vol.~43, no.~1, pp. 44--55, 1996.

\bibitem{ODonnell1988a}
M.~O'Donnell and S.~W. Flax, ``{Phase-Aberration Correction Using Signals From
  Point Reflectors and Diffuse Scatterers: Basic Principales},'' \emph{IEEE
  Transactions on Ultrasonics, Ferroelectrics, and Frequency Control}, vol.~35,
  no.~6, pp. 768--774, 1988.

\bibitem{Liu1994}
D.~Liu and R.~C. Waag, ``Correction of ultrasonic wavefront distortion using
  backpropagation and a reference waveform method for time-shift
  compensation.'' \emph{The Journal of the Acoustical Society of America}, vol.
  96 2 Pt 1, pp. 649--60, 1994.

\bibitem{Ng1994}
\BIBentryALTinterwordspacing
G.~Ng, S.~Worrell, P.~Freiburger, and G.~Trahey, ``{A comparative evaluation of
  several algorithms for phase aberration correction},'' \emph{IEEE
  Transactions on Ultrasonics, Ferroelectrics and Frequency Control}, vol.~41,
  no.~5, pp. 631--643, 1994. [Online]. Available:
  \url{http://ieeexplore.ieee.org/lpdocs/epic03/wrapper.htm?arnumber=308498}
\BIBentrySTDinterwordspacing

\bibitem{Nock1989}
L.~Nock, G.~E. Trahey, and S.~W. Smith, ``{Phase aberration correction in
  medical ultrasound using speckle brightness as a quality factor},'' \emph{The
  Journal of the Acoustical Society of America}, vol.~85, no.~5, pp.
  1819--1833, 1989.

\bibitem{Fink1992}
M.~Fink, ``Time reversal of ultrasonic fields—part i: Basic principles,''
  \emph{IEEE Transactions on Ultrasonics, Ferroelectrics, and Frequency
  Control}, vol.~39, no.~5, pp. 555--566, 1992.

\bibitem{Karaman1993}
M.~Karaman, A.~Atalar, H.~Koymen, and M.~O'Donnell, ``{A phase aberration
  correction method for ultrasound imaging.}'' \emph{IEEE transactions on
  ultrasonics, ferroelectrics, and frequency control}, vol.~40, no.~4, pp.
  275--282, 1993.

\bibitem{Byram2015}
B.~Byram, K.~Dei, J.~Tierney, and D.~Dumont, ``{A model and regularization
  scheme for ultrasonic beamforming clutter reduction},'' \emph{IEEE
  Transactions on Ultrasonics, Ferroelectrics, and Frequency Control}, vol.~62,
  no.~11, pp. 1913--1927, 2015.

\bibitem{Shaswary2015}
E.~Shaswary, J.~Tavakkoli, and Y.~Xu, ``{A new algorithm for time-delay
  estimation in ultrasonic echo signals [Correspondence]},'' \emph{IEEE
  Transactions on Ultrasonics, Ferroelectrics, and Frequency Control}, vol.~62,
  no.~1, pp. 236--241, 2015.

\bibitem{Viola2003}
F.~Viola and W.~F. Walker, ``{A comparison of the performance of time-delay
  estimators in medical ultrasound},'' \emph{IEEE Transactions on Ultrasonics,
  Ferroelectrics, and Frequency Control}, vol.~50, no.~4, pp. 392--401, 2003.

\bibitem{Byram2012}
B.~C. Byram, G.~E. Trahey, and J.~A. Jensen, ``{A method for direct localized
  sound speed estimates using registered virtual detectors},'' \emph{Ultrasonic
  Imaging}, vol.~34, no.~3, pp. 159--180, 2012.

\bibitem{Anderson1998}
\BIBentryALTinterwordspacing
M.~E. Anderson and G.~E. Trahey, ``{The direct estimation of sound speed using
  pulse–echo ultrasound},'' \emph{The Journal of the Acoustical Society of
  America}, vol. 104, no.~5, pp. 3099--3106, 1998. [Online]. Available:
  \url{http://asa.scitation.org/doi/10.1121/1.423889}
\BIBentrySTDinterwordspacing

\bibitem{Jakovljevic2018}
\BIBentryALTinterwordspacing
M.~Jakovljevic, S.~Hsieh, R.~Ali, G.~{Chau Loo Kung}, D.~Hyun, and J.~J. Dahl,
  ``{Local speed of sound estimation in tissue using pulse-echo ultrasound:
  Model-based approach},'' \emph{The Journal of the Acoustical Society of
  America}, vol. 144, no.~1, pp. 254--266, 2018. [Online]. Available:
  \url{http://asa.scitation.org/doi/10.1121/1.5043402}
\BIBentrySTDinterwordspacing

\bibitem{Slack2014}
T.~Slack, C.~Myers, C.~Martin, and S.~Heymsfield, ``The geographic
  concentration of us adult obesity prevalence and associated social, economic,
  and environmental factors,'' \emph{Obesity (Silver Spring, Md.)}, vol.~22,
  March 2014.

\bibitem{Alm2008}
A.-M. Almeida, H.-P. Cotrim, D.-B. Barbosa \emph{et~al.}, ``Fatty liver disease
  in severe obese patients: Diagnostic value of abdominal ultrasound,''
  \emph{World Journal of Gastroenterology}, vol.~14, p. 1415, 03 2008.

\bibitem{Dahl2003}
J.~Dahl and G.~Trahey, ``Off-axis scatterer filters for improved aberration
  measurements,'' vol.~2, 11 2003, pp. 343-- 347 Vol.1.

\bibitem{Dahl2008}
J.~Dahl and T.~Feehan, ``Direction of arrival filters for improved aberration
  estimation,'' \emph{Ultrasonic imaging}, vol.~30, pp. 1--20, 02 2008.

\bibitem{Shin2018}
J.~Shin, L.~Huang, and J.~Yen, ``Spatial prediction filtering for medical
  ultrasound in aberration and random noise,'' \emph{IEEE Transactions on
  Ultrasonics, Ferroelectrics, and Frequency Control}, vol.~PP, pp. 1--1, Oct
  2018.

\bibitem{Dahl2017}
J.~Dahl, D.~Hyun, Y.~Li, M.~Jakovljevic, M.~Lediju~Bell, W.~Long, N.~Bottenus,
  V.~Kakkad, and G.~Trahey, ``Coherence beamforming and its applications to the
  difficult-to-image patient,'' Sep 2017, pp. 1--10.

\bibitem{Gong2015}
P.~{Gong}, M.~C. {Kolios}, and Y.~{Xu}, ``Delay-encoded transmission and image
  reconstruction method in synthetic transmit aperture imaging,'' \emph{IEEE
  Transactions on Ultrasonics, Ferroelectrics, and Frequency Control}, vol.~62,
  no.~10, pp. 1745--1756, 2015.

\bibitem{JENSEN2006}
J.~A. Jensen, S.~I. Nikolov, K.~L. Gammelmark, and M.~H. Pedersen, ``Synthetic
  aperture ultrasound imaging,'' \emph{Ultrasonics}, vol.~44, pp. e5 -- e15,
  2006.

\bibitem{Dahl2005}
J.~J. Dahl, D.~A. Guenther, and G.~E. Trahey, ``{Adaptive Imaging and Spatial
  Compounding in the Presence of Aberration},'' \emph{IEEE Transactions on
  Ultrasonics, Ferroelectrics, and Frequency Control}, vol.~52, no.~7, pp.
  1131--1144, 2005.

\bibitem{Goss1978}
S.~A. Goss, R.~L. Johnston, and F.~Dunn, ``{Comprehensive compilation of
  empirical ultrasonic properties of mammalian tissues},'' \emph{Journal of the
  Acoustical Society of America}, vol.~64, no.~2, pp. 423--457, 1978.

\bibitem{Mallart1991}
R.~Mallart and M.~Fink, ``{The van Cittert-Zernike theorem in pulse echo
  measurements},'' \emph{Journal of the Acoustical Society of America},
  vol.~90, no.~5, pp. 2718--2727, 1991.

\bibitem{kak2002}
A.~C. Kak, M.~Slaney, and G.~Wang, ``Principles of computerized tomographic
  imaging,'' \emph{Medical Physics}, vol.~29, no.~1, pp. 107--107, 2002.

\end{thebibliography}

\onecolumn

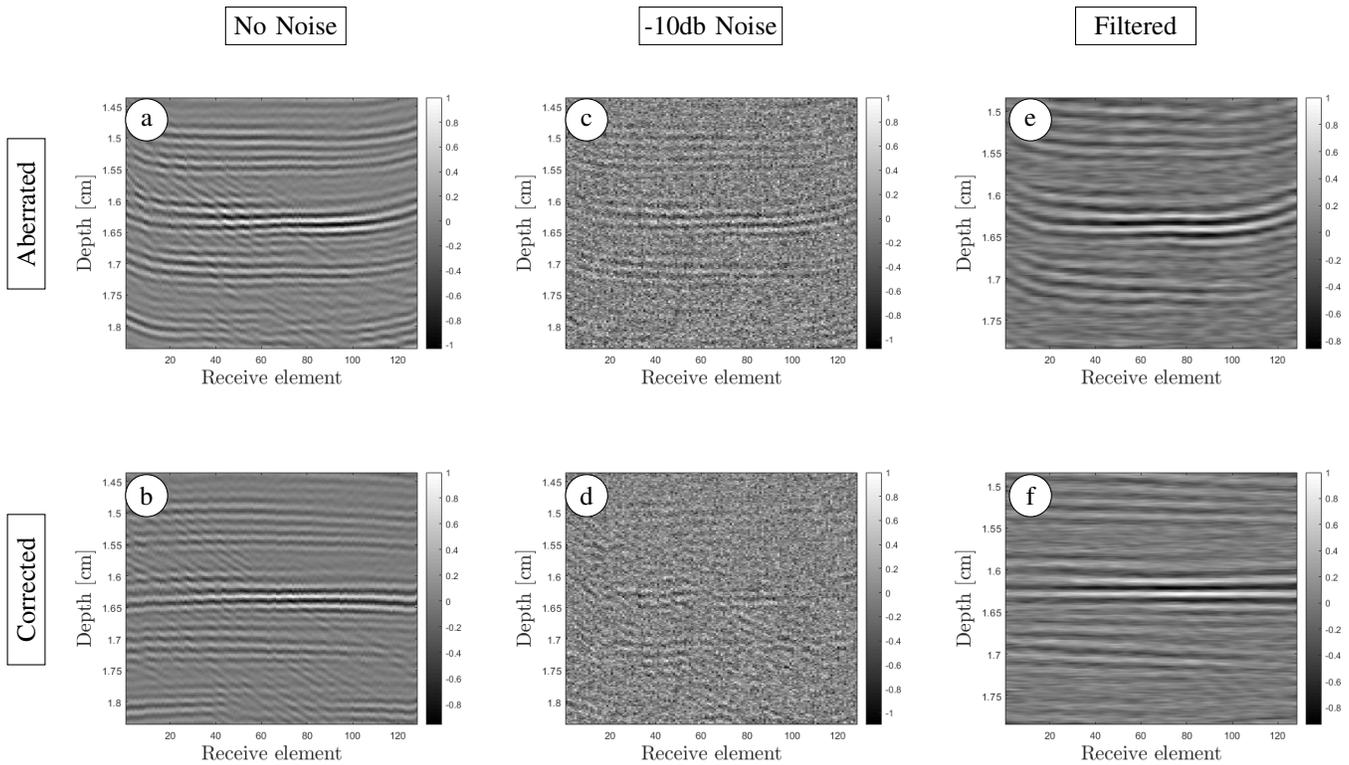
\begin{figure}
\centering
\begin{tikzpicture}

\node[inner sep=0pt] (org) at (0,0)
    {\includegraphics[width=.31\textwidth]{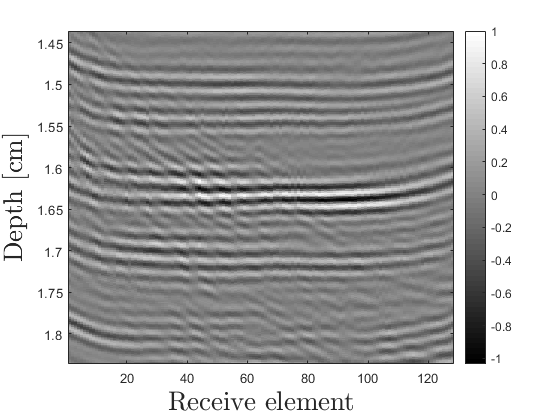}};
    \draw [fill=white](-0.8,2.5) rectangle (0.8,3) node[pos=.5] {No Noise};
   \draw [fill=white](-3.2,-0.75) rectangle (-3.7,1.25) node[pos=.5, rotate=90 ] {Aberrated};
      \draw[fill=white] (-1.85,1.5) circle (0.28cm) node[text=black] {a};
 
    \node[inner sep=0pt] (filter) at (0,-5)
    {\includegraphics[width=.31\textwidth]{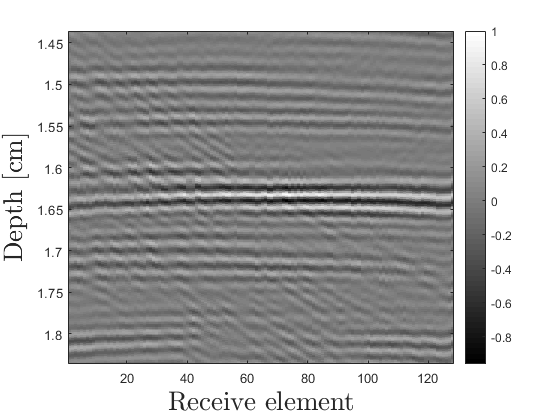}};
      \draw [fill=white](-3.2,-5.75) rectangle (-3.7,-3.75) node[pos=.5, rotate=90 ] {Corrected};
      \draw[fill=white] (-1.85,-3.5) circle (0.28cm) node[text=black] {b};
 
 \node[inner sep=0pt] (fft) at (5.85,0)
    {\includegraphics[width=.31\textwidth]{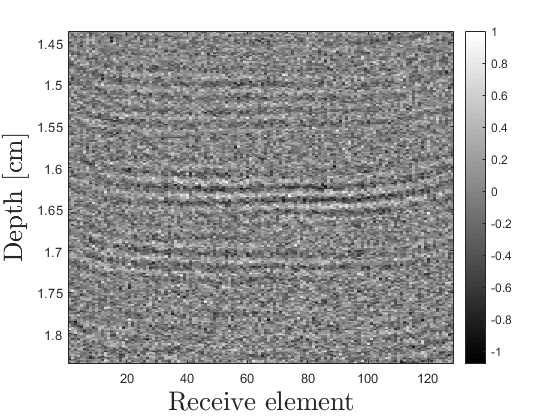}};
   \draw [fill=white](4.7,2.5) rectangle (6.6,3) node[pos=.5] {-10db Noise};
      \draw[fill=white] (4,1.5) circle (0.28cm) node[text=black] {c};

     \node[inner sep=0pt] (fftfilt) at (5.85,-5)
      {\includegraphics[width=.31\textwidth]{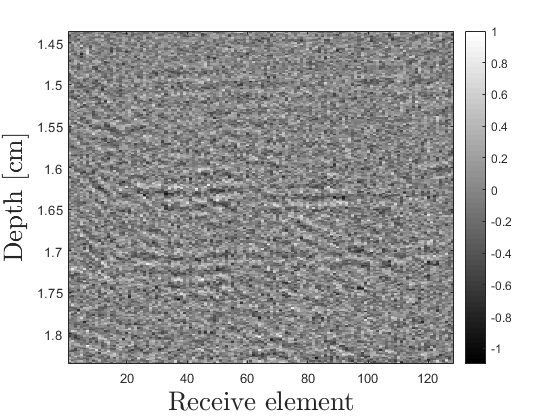}};
      \draw[fill=white] (4,-3.5) circle (0.28cm) node[text=black] {d};

\node[inner sep=0pt] (fft) at (11.7,0)
    {\includegraphics[width=.31\textwidth]{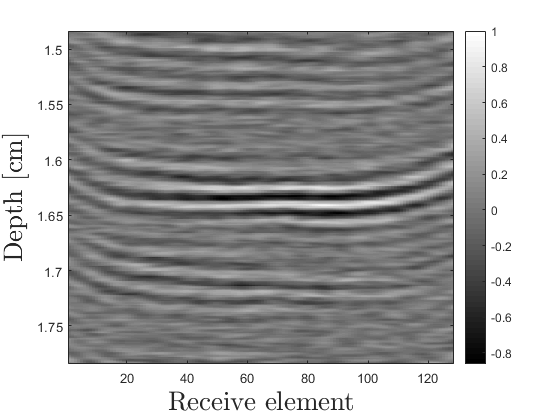}};
   \draw [fill=white](10.5,2.5) rectangle (12.1,3) node[pos=.5] {Filtered};
    \draw[fill=white] (9.9,1.5) circle (0.28cm) node[text=black] {e};     
    
    \node[inner sep=0pt] (filt) at (11.7,-5)
         {\includegraphics[width=.31\textwidth]{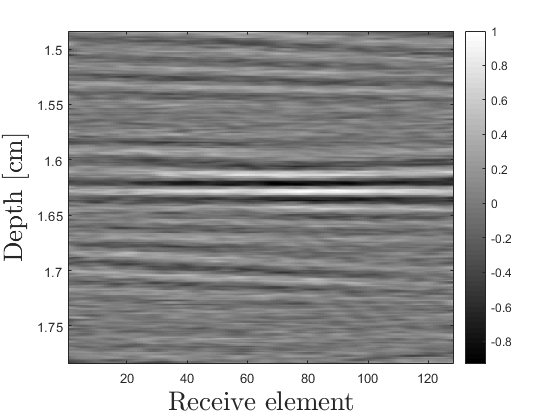}};
      \draw[fill=white] (9.9,-3.5) circle (0.28cm) node[text=black] {f};
    
\end{tikzpicture}
   \caption{  Plots of Transmit beamformed aberrated RF signals  (a) with no noise, (c) with added noise of -10dB SNR and (e) after filtration of the noisy signals in (c). (b, d) and (f) show their corresponding plots after phase aberration correction with NCC.  Focal point was at a speckle point at z=1.63 cm. }
   \label{fig:sub6}
\end{figure}

\begin{figure}
\centering
   \includegraphics[width=.34\textwidth]{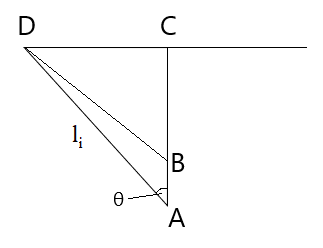}
   \caption{The illustration of truth travel delay calculation. Points D shows the transducer element and point A is the focal point position. C is the transducer element right about the focal point A }
   \label{fig:sub12}
\end{figure}

\begin{table}
\begin{center}
\caption{Quantitative comparison of aberrated and NCC and f-NCC corrected images of simulated phantom}
\begin{tabular}{@{}ccc@{}}
\toprule
          & PSNR {[}dB{]} & CNR hyper {[}dB{]} \\ \midrule
Aberrated & 5.4          & 1.54               \\
NCC & 5.8          & 1.2               \\
f-NCC & 8.2          & 1.92               \\ \bottomrule
\end{tabular}
\label{tab1}
 \end{center}
\end{table}

\end{document}